\documentclass[twocolumn,aps,
prl,reprint,
superscriptaddress,
amsmath,amssymb,
nofootinbib]{revtex4-2}
\renewcommand{\selectlanguage}[1]{}

\usepackage[pdftex]{graphicx,color}
\usepackage[pdftex,
            colorlinks=true,
            citecolor=blue,
            linkcolor=blue]{hyperref}
\graphicspath{{./figure},{./fig_SM}}

\usepackage{braket}
\usepackage{times,multirow,amsfonts,bm,xspace,pifont}
\usepackage{soul}
\usepackage[normalem]{ulem}
\usepackage{pdfpages}

\usepackage{hyperref}
\hypersetup{% hyperrefオプションリスト
 setpagesize=false,%
 colorlinks=true,%
 linkcolor=blue,
 citecolor=blue,
}

\makeatletter
\AtBeginDocument{\let\LS@rot\@undefined}
\makeatother

\begin{document}

\newcommand*\YY[1]{\textcolor{blue}{#1}}
\newcommand{\YYS}[1]{\textcolor{blue}{\sout{#1}}}

\newcommand*\HL[1]{\textcolor{black}{#1}}%{red}{#1}}

\title{Odd-parity magnetism by quantum geometry}

\author{Kanta Kudo}
\email[]{kudo.kanta.37u@st.kyoto-u.ac.jp}
\affiliation{Department of Physics, Graduate School of Science, Kyoto University, Kyoto 606-8502, Japan}

\author{Youichi Yanase}
\affiliation{Department of Physics, Graduate School of Science, Kyoto University, Kyoto 606-8502, Japan}
\date{\today}

\begin{abstract}
We uncover a geometric mechanism of \HL{odd-parity} multipole magnetism driven by the \HL{quantum metric} of Bloch electrons.
%that is classified as odd-parity multipole magnetism. 
By analyzing spin and odd-parity multipole susceptibilities in a multi-sublattice model, we demonstrate that %the ferroic multipole fluctuation is induced by
the \HL{quantum metric} directly controls the instability toward odd-parity magnetic multipole order over a wide range of parameters, %Due to the Hubbard interaction, %the quantum-geometric multipole fluctuations 
which condenses under Hubbard interaction. The resulting state exhibits  complex magnetic correlations, as a hallmark of quantum-geometric magnetism. \HL{These results establish a geometric design principle for odd-parity multipole magnets and provide a route toward the experimental verification of quantum-geometric magnetism.} 
\end{abstract}

\maketitle

\textit{Introduction} ---
The Bloch wave function that describes the electronic states in solids varies with crystal momentum, giving rise to geometric quantities such as the quantum geometric tensor that characterizes the geometric structure of Bloch states~\cite{xiao2010berry,resta2011insulating}. The imaginary part of the quantum geometric tensor, the Berry curvature, and the real part, the quantum metric, play central roles in a wide range of phenomena in quantum materials~\cite{xiao2010berry,Torma2022,Torma2023,Rossi2021,Morimoto_JPSJreview,Nagaosa-Yanase,Yu2024,resta2011insulating}. %Various phenomena are related to the Berry curvature, quantum metric, and other geometric quantities, as demonstrated by intensive studies in condensed matter physics~\cite{xiao2010berry,Torma2022,Torma2023,Rossi2021,Morimoto_JPSJreview,Nagaosa-Yanase,Yu2024,resta2011insulating}. 
    
While the Berry curvature underlies various Hall responses~\cite{TKNN,KOHMOTO1985343,Haldane1988,xiao2010berry,sodemann2015quantum},
the quantum metric has been identified as a key ingredient of superfluid density~\cite{peotta2015superfluidity,liang2017band,Torma2022,Rossi2021,Torma2023,Tian2023,Tanaka2025-sy,Banerjee2025-ew}, nonlinear optical response~\cite{watanabe2021chiral,Morimoto_JPSJreview,Nagaosa-Yanase,Ahn2020} and light-matter interaction~\cite{topp2021light}, superconducting diode effect~\cite{Hu2025}, and anomalous Landau levels~\cite{Rhim2021}. More recently, quantum geometry has also been shown to influence the stability of correlated many-body states, including ferromagnetism and unconventional superconductivity~\cite{kitamura2024spin,heinsdorf2025altermagnetic,Yu2024,Wu2020,Kang2024,Han2024,Alicea2025,jahin2025enhancedkohnluttingertopologicalsuperconductivity,kitamura2025quantumgeometricferromagnetismsingular,Kitamura_anapole,Gassner2024}. 
%For example, quantum geometry can induce ferromagnetism and spin-triplet superconductivity through the quantum metric~\cite{kitamura2024spin}.
Despite these advances, the interplay between quantum geometry and electron correlations remains largely elusive. In particular, it is unclear whether quantum geometry plays a fundamental role beyond entering as a form factor in conventional susceptibility analysis. Furthermore, signatures of magnetism with geometric origin have yet to be explored. These issues are particularly acute for unconventional magnetic orders with complex internal structures.

In parallel, unconventional magnetism that breaks time-reversal symmetry with vanishing spontaneous magnetization has attracted considerable attention. 
Such magnetism is classified into higher-order multipole magnetism~\cite{watanabe2018group,Hayami2018} including altermagnetism~\cite{Altermagnetism_PRX,Naka2019} that produces non-relativistic spin splitting even in the absence of spin-orbit coupling. Unconventional magnetism causes nontrivial symmetry breaking and allows emergent electromagnetic responses including the magnetoelectric and Edelstein effects~\cite{Curie,EDELSTEIN1990233}, longitudinal spin current generation~\cite{Altermagnetism_PRX,Naka2019}, magnetopiezoelectric effect~\cite{watanabe2017magnetic,Shiomi2019}, and various nonreciprocal responses~\cite{Morimoto_JPSJreview,Nagaosa-Yanase,watanabe2021chiral,Ahn2020,Gao2014,watanabe2020}. 
%which have been intensively studied for potential application. 
In addition, multipole fluctuations can produce unconventional superconductivity~\cite{Kozii-Fu2015,Sumita2020}. 
In light of the recent surge of interest, establishing a design principle for unconventional magnets is highly anticipated.  
%and quantum geometry, the interrelation of these two disciplines has emerged as a fascinating subject.  

%The effects of quantum geometry on unconventional magnetic ordering have not been studied except in recent work on altermagnetism~\cite{heinsdorf2025altermagnetic}.

In this Letter, we uncover a quantum-geometric mechanism of odd-parity multipole magnetism driven by the quantum metric. We demonstrate that the quantum metric directly controls the ferroic instability toward odd-parity magnetic multipole order over a wide parameter range. Our approach successfully reveals the quantum-metric contribution to unconventional magnetism with multi-sublattice structure in a direct and nontrivial manner. 
By analyzing spin and multipole susceptibilities, we predict complex magnetic correlations as a hallmark of quantum-geometric magnetism, which can be verified experimentally.
%We also show complex magnetic correlations, which are a hallmark of quantum-geometric magnetism and can be verified by experiments.

    \begin{figure}[tbp]
      %\centering
      %\begin{subfigure}{0.55\linewidth}
%        \centering
        %\caption{Top view}       \includegraphics[width=\linewidth]{intralayer_hopping.png}
%        \caption{.}
        %\label{fig:left}
      %\end{subfigure}
      %\hfill 
      %\begin{subfigure}{0.35\linewidth}
%        \centering
        %\caption{Side view} \includegraphics[width=\linewidth]%{interlayer_hopping.png}
%        \caption{}
        %\label{fig:right}
      %\end{subfigure}
      \includegraphics[width=8cm]{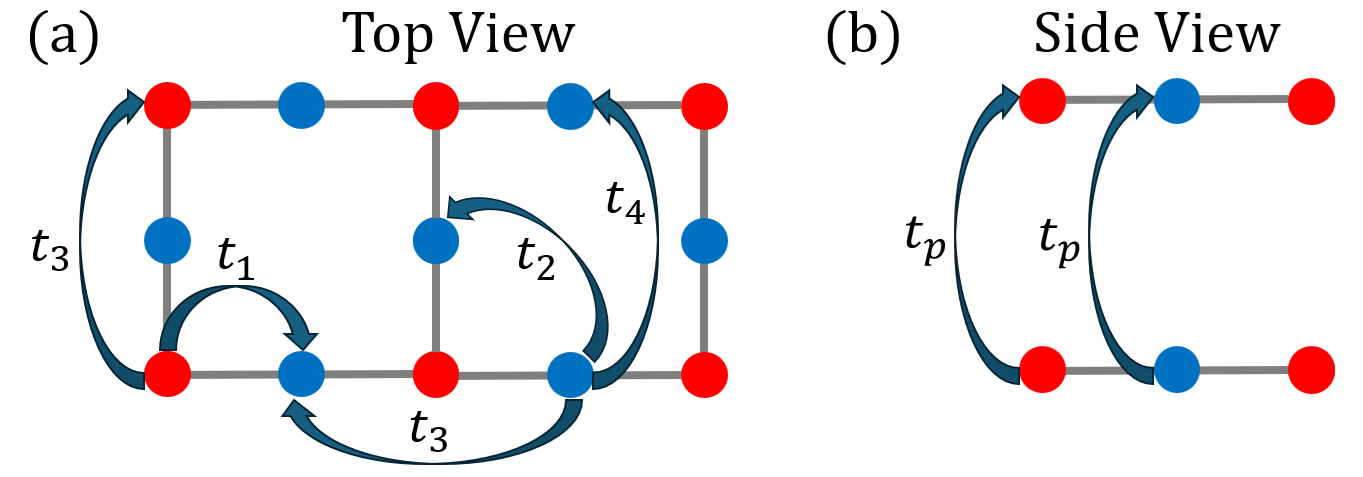}
      \caption{Lattice structure of the bilayer Lieb lattice model. (a) Top view illustrating intra-layer hopping parameters with arrows. (b) Side view and the inter-layer hopping parameter. %Each arrow shows intra and inter layer hopping in bilayer Lieb lattice.
      }
    \label{hopping}
    \end{figure}

\textit{Model} ---
    \HL{%To study the standard ferromagnetism and odd-parity magnetism on equal footing, 
    In the following, we analyze the bilayer Lieb lattice model. We choose the Lieb lattice because it has been widely adopted as a standard model in theoretical studies of quantum geometry~\cite{kitamura2024spin,Torma2022,Torma2023,peotta2015superfluidity,julku2016geometric, huhtinen2022revisiting,chau2026quantum,xiao2025effects}. Significant quantum-geometric effects emerge, and their characteristics are expected to be generic beyond the Lieb lattice~\cite{liang2017band,Tian2023,Tanaka2025-sy,Banerjee2025-ew,kitamura2025quantumgeometricferromagnetismsingular}. The bilayer structure is introduced to the model to study the standard ferromagnetism and odd-parity magnetism on equal footing.} %to have strong quantum-geometric effect. 
    %Moreover, bilayer Lieb lattice is predicted to show one typical quantum-geometric effect in untwisted bilayer system which single layer has degenerate point (see Appendix A in Supplemental Material \cite{SM}).} Figure~\ref{hopping} shows the lattice structure and the hopping integrals involved in the model. %intra- and inter-layer hopping in this model. Thus, 
    There are three sublattices $j=A, B, C$ in each layer $l=1,2$. Spin magnetization and spin multipole moment in the unit cell are defined by $\hat{M}_{\rm E} = \sum_{j} (\hat{S}_{j1} + \hat{S}_{j2})$ and $\hat{M}_{\rm O} = \sum_{j} (\hat{S}_{j1} - \hat{S}_{j2})$, respectively. In this way, the spin multipole moment can arise from the layer degree of freedom~\cite{watanabe2018group,Hayami2018}. The spin multipole moment is an odd-parity order parameter that breaks not only time-reversal symmetry but also space inversion symmetry.
    
    The single-particle Bloch Hamiltonian of the model is written as 
    \begin{align}
        \hat{H}_0(\bm{k}) = \begin{pmatrix}
                    \hat{H}_{\rm intra}(\bm{k}) & \hat{H}_{\rm inter}(\bm{k})\\
                    \hat{H}_{\rm inter}(\bm{k}) & \hat{H}_{\rm intra}(\bm{k})\\
                  \end{pmatrix} \otimes \hat{\sigma}_0,
                  \label{eq:hamiltonian}
    \end{align}
    where $\hat{H}_{\rm intra}$ and $\hat{H}_{\rm inter}$ denote intra- and inter-layer hopping Hamiltonian, respectively. We consider systems without spin-orbit coupling, and thus the Hamiltonian is proportional to $\hat{\sigma}_0$ in spin space. The matrix elements are
    %\YY{
    \begin{align}
        (\hat{H}_{\rm intra}(\bm{k}))_{11} &= -2t_3(\cos k_x + \cos k_y),\\
        (\hat{H}_{\rm intra}(\bm{k}))_{12} &= -2t_1\cos \frac{k_x}{2},\\
        (\hat{H}_{\rm intra}(\bm{k}))_{13} &= -2t_1\cos \frac{k_y}{2},\\
        (\hat{H}_{\rm intra}(\bm{k}))_{22} &= -2t_3\cos k_x -2t_4\cos k_y,\\
        (\hat{H}_{\rm intra}(\bm{k}))_{23} &= -4t_2\cos \frac{k_x}{2}\cos \frac{k_y}{2},\\
        (\hat{H}_{\rm intra}(\bm{k}))_{33} &= -2t_3\cos k_y -2t_4\cos k_x,\\
        \hat{H}_{\rm inter}(\bm{k}) &= -t_{\rm p} \hat{I}_{3\times 3},
    \end{align}
    %}
    where $t_1$, $t_2$, $t_3$, $t_4$ are the intra-layer hopping parameters and $t_{\rm p}$ is the inter-layer one, as illustrated in Fig.~\ref{hopping}.  
    $I_{3\times 3}$ is the $3\times 3$ unit matrix. In the following, we set $t_1  =1$ as the unit of energy. The total Hamiltonian is $\hat{H} = \hat{H}_{0} + \hat{H}_{\rm I}$ including the Hubbard interaction term $\hat{H}_{\rm I} = U \sum_{i} \hat{n}_{i\uparrow} \hat{n}_{i\downarrow}$, where $i$ labels the unit cell and the sublattice. The coupling constant $U$ represents the on-site Coulomb interaction. %and $N$ is the number of unit cells.

\begin{figure}[tbp]
%        \centering
        \includegraphics[width=7cm]{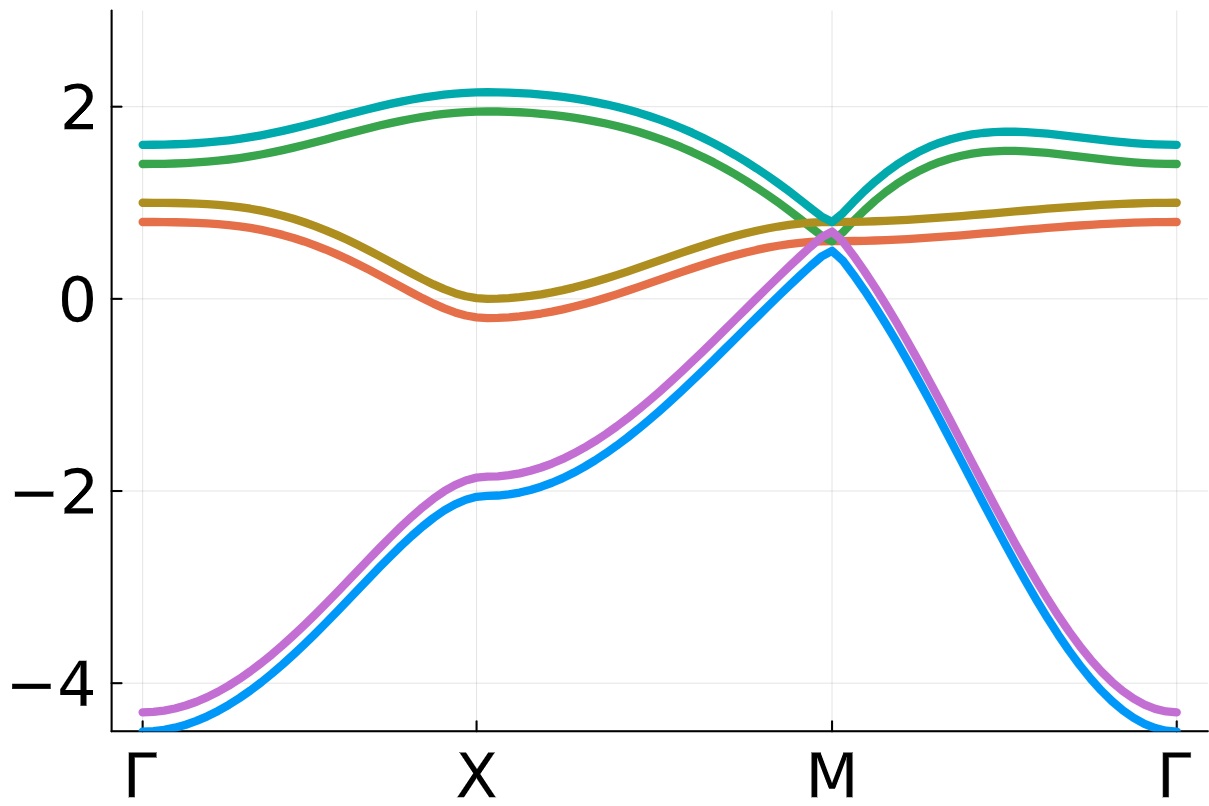}
        \caption{Band dispersion of the bilayer Lieb lattice model for $(t_1,t_2,t_3,t_4,t_{\rm p})=(1, 0.4, 0.15, 0.2, 0.1)$.}
        \label{energy}
    \end{figure}

Because the single-particle Hamiltonian commutes interlayer mirror operation $\hat{\sigma}_{\rm{h}}$, mirror parity $p=\pm 1$ is a good quantum number that labels Bloch states. Therefore, the electron bands are distinguished by the band index $n=1,2,3$ and the mirror parity $p = \pm 1$. This feature enables a transparent comparison of spin and multipole fluctuations, as shown later. 
Figure~\ref{energy} shows the energy dispersion $\epsilon_n^{p}(\bm{k})$ for the parameter set $(t_1,t_2,t_3,t_4,t_{\rm p})=(1, 0.4, 0.15, 0.2,0.1)$, which is adopted in the following part. \HL{The following results remain unchanged for other choices of parameters and also for variations of the model, as shown in Section~A of the Supplemental Material~\cite{SM}.} The quantum numbers of the bands are $(n,p) = (1,+), (1,-), (2,+), (2,-), (3,+), (3,-)$ from bottom to top, except around the M point. %where the bands cross.
Electronic states with the same mirror parity degenerate at the M point, as the band degeneracy is protected by $D_{4h}$ point group symmetry~\cite{kitamura2024spin}.
    %bands with the same mirror eigenvalue at energies 0.6 and 0.9 are degenerate. 
In addition, bands with different mirror parity cross around the M point without a gap opening. %at energies 0.6 and 0.9 are degenerate.

\textit{\HL{Roles of quantum geometry}} --- 
    First, we discuss the bare spin susceptibility $\chi^{\rm{E}}_0(\bm{q})$ of even-parity and the odd-parity multipole susceptibility $\chi^{\rm{O}}_0(\bm{q})$, which are defined by
    \begin{align}
        \chi^{\rm{E/O}}_0(\bm{q}) = i %\frac{i}{\hbar} 
        \int^{\infty}_{0}\langle[\hat{M}_{\rm{E/O}}(\bm{q},t),\hat{M}_{\rm{E/O}}(-\bm{q})]\rangle_0 e^{-\delta t} dt.
    \end{align}
    Here, $\hat{M}_{\rm{E}}(\bm{q})$ and $\hat{M}_{\rm{O}}(\bm{q})$ are the operators of spin magnetization and spin multipole moment at the wave vector $\bm{q}$, respectively (see Section~B in Supplemental Material~\cite{SM}). The expectation values $\langle\hat{A}\rangle_0$ %represents expectation value of quantity $\hat{A}$ obtained 
    are calculated based on the single-particle Hamiltonian.
    The momentum dependence and relative strength of $\chi^{\rm{E}}_0(\bm{q})$ and $\chi^{\rm{O}}_0(\bm{q})$ capture long-range magnetic order triggered by the Coulomb interaction~\cite{Nogaki2024} and the critical fluctuation.
    \HL{The momentum ${\bm q}$ for the maximum of susceptibilities indicates spin and multipole correlations.}
    \HL{Since quantum metric has been shown to enhance ferromagnetic spin correlations~\cite{kitamura2024spin}, we can expect further diverse ferroic fluctuations~\cite{heinsdorf2025altermagnetic} to arise from quantum geometry.}

    %The contrasting behaviors of spin and multipole fluctuations can be understood by analyzing the curvature of susceptibilities. 
    The strength of ferroic fluctuations relative to antiferroic fluctuations can be quantified by the curvature at $\bm{q}= 0$~\cite{kitamura2024spin,heinsdorf2025altermagnetic,kitamura2025quantumgeometricferromagnetismsingular}. When a ferroic fluctuation develops, the susceptibility shows a strong peak at ${\bm q}=0$, and the curvature becomes largely negative. However, positive curvature indicates that susceptibility has a local minimum at ${\bm q}=0$, ruling out ferroic fluctuation. 
    %    If ferro magnetic/multipole fluctuation exists, the curvature of susceptibility must be minus.  
    %    Due to 4-fold rotational symmetry, hessian matrix is diagonal and each eigenvalue is equal. So, it is enough to consider one diagonal element of hessian matrix to discuss curvature and hereafter we call one diagonal element of hessian matrix curvature. 
%
    \HL{The following analysis of the curvatures reveals the intimate relation between the spin and multipole fluctuations and quantum geometry.} %the contrasting roles of quantum geometry in spin and multipole orders.}

     \HL{Since mirror parity is a good quantum number, the overlap of wave functions with different mirror parity vanishes. Therefore, the quantum geometric tensor %that includes the quantum metric and the Berry curvature 
     can be defined within the mirror sector. This property highlights the difference between the quantum-geometric effects on spin susceptibility and those on multipole susceptibility. In particular, we demonstrate that the quantum metric, defined as ${g^{ij}_{n}}(\bm{k}) = {\rm Re} \left(\bra{\partial_i u_n^p(\bm{k})} (1-\ket{u_n^{p}(\bm{k})}\bra{u_n^{p}(\bm{k})})\ket{\partial_j u_n^p(\bm{k})}\right)$ (see Section~C in Supplemental Material~\cite{SM}), plays essential but distinct roles in spin and multipole fluctuations.} 
     
    Because of four-fold rotation symmetry of the system, the curvature of spin and multipole susceptibilities is given by $\partial_{q_x}^2\chi^{\rm{E/O}}_0(\bm{q})\big|_{{\bm q}=0}$, which can be expressed by 
    \begin{align}
    &\partial_{q_i}\partial_{q_j}\chi^{\rm{E}}_0(\bm{q})\big|_{{\bm q}=0} = %\notag \\
        %& 
        \int\frac{d\bm{k}}{(2\pi)^2} \Big[  \sum_{n,p}\frac{f^{\prime\prime}(\epsilon_n^p(\bm{k}))}{6}\partial_i\partial_j\epsilon_n^{p}(\bm{k}) \notag \\
        &- \sum_{n,m,p}L_{nm}^{pp}(\bm{k},\bm{0}) \{\delta_{nm}g_n^{ij}(\bm{k}) - (1-\delta_{nm})g_{nm}^{ij}(\bm{k}) \} \Big],
    \label{eq:curvature_spin}
    \end{align}
    and 
    \begin{align}
        &\partial_{q_i}\partial_{q_j}\chi^{\rm{O}}_0(\bm{q})\big|_{{\bm q}=0} = \notag \\
        & \int\frac{d\bm{k}}{(2\pi)^2} \Big[ \sum_{n,p} \frac{1}{2} \Big\{\frac{\partial_i\epsilon_n^{p}f^{\prime}(\epsilon_n^{p}) + \partial_i\epsilon_n^{-p}f^{\prime}(\epsilon_n^{-p})}{2}\frac{\partial_j\epsilon_n^{p} + \partial_j\epsilon_n^{-p}}{(\epsilon_n^p-\epsilon_n^{-p})^2} \notag \\ &+ \frac{\partial_i\epsilon_n^{-p}\partial_j\epsilon_n^{-p}f^{\prime\prime}(\epsilon_n^{-p}) - \partial_i\epsilon_n^{p}\partial_j\epsilon_n^{p}f^{\prime\prime}(\epsilon_n^{p})}{4}\frac{1}{\epsilon_n^p-\epsilon_n^{-p}} \notag \\ &+ \frac{f(\epsilon_n^{-p})-f(\epsilon_n^{p})}{2}\frac{(\partial_i\epsilon_n^{p} + \partial_i\epsilon_n^{-p})(\partial_j\epsilon_n^{p} + \partial_j\epsilon_n^{-p})}{(\epsilon_n^p-\epsilon_n^{-p})^3} \Big\} \notag \\ 
        &- \sum_{n,m,p} L_{nm}^{p-p}(\bm{k},\bm{0}) \{\delta_{nm} g_n^{ij}(\bm{k}) - (1-\delta_{nm}) g_{nm}^{ij}(\bm{k})\} \Big],
        \label{eq:curvature_multipole}
    \end{align}
    with the Lindhard function $L_{nm}^{pp^{\prime}}(\bm{k},\bm{q})$ defined as 
    \begin{align}
        & L_{nm}^{pp^{\prime}}(\bm{k},\bm{q}) = \frac{f(\epsilon^{p^{\prime}}_m({\bm k})) - f(\epsilon_n^p({\bm k}+{\bm q}))}{\epsilon_n^p({\bm k}+{\bm q}) - \epsilon^{p^{\prime}}_m({\bm k})}.
    \end{align}
    Here, $f(\epsilon)$ is the Fermi distribution function, and we set the temperature $T=0.01$ throughout this Letter. We denote $\epsilon_n^{p}({\bm k})$ by $\epsilon_n^{p}$ in Eq.~\eqref{eq:curvature_multipole} for simplicity.
% 
% Since mirror parity is a good quantum number, the overlap of wave functions with different mirror parity vanishes. Therefore the quantum geometric tensor that includes the quantum metric and the Berry curvature can be defined within the mirror sector. This property will highlight the difference between the quantum-geometric effects on spin susceptibility and multipole susceptibility(see Appendix C in Supplemental Material \cite{SM}). 
% 
 While the first terms of the curvatures %these equations 
 contain only the effects of band dispersion, the second terms represent the quantum-geometric effects, which are determined by the quantum metric $g_n^{ij}(\bm{k})$ and the band-resolved quantum metric $g_{nm}^{ij}(\bm{k})$ in the mirror sector. Thus, we call the first terms in Eqs.~\eqref{eq:curvature_spin} and \eqref{eq:curvature_multipole} band-dispersion terms and the second terms geometric terms.

    Interestingly, the geometric terms for spin susceptibility and multipole susceptibility are expressed by almost the same formula. The difference is only the sign of $\pm p$ in $L_{nm}^{p \pm p}(\bm{k},\bm{0})$. Using this simple relationship, we can discuss the contrasting roles of quantum geometry in spin and multipole correlations coherently. The significant difference arises from the quantum metric term proportional to $g_n^{ij}(\bm{k})$. This term in Eqs.~\eqref{eq:curvature_spin} and \eqref{eq:curvature_multipole} is always negative and thus favors ferroic fluctuations~\cite{kitamura2024spin}. The multiplied factor is $L_{nn}^{p p}(\bm{k},\bm{0})=-f'(\epsilon_n^{p}({\bm k}))$ for spin susceptibility, which is reduced to the derivative of the Fermi distribution function.
    Therefore, the quantum metric term that favors ferromagnetic fluctuations is a Fermi surface term, and it can be enhanced when the Fermi level lies on band degeneracy points. In contrast, the quantum metric term for multipole susceptibility is proportional to $L_{nn}^{p -p}(\bm{k},\bm{0}) = (f(\epsilon^{-p}_n({\bm k})) - f(\epsilon_n^p({\bm k})))/ (\epsilon_n^p({\bm k}) - \epsilon^{-p}_n({\bm k}))$, and thus is an inter-band term resulting from inter-layer band splitting. Therefore, the quantum metric can enhance ferroic multipole fluctuations even when the band degeneracy points are away from the Fermi level, provided that the Fermi level lies within the inter-layer band splitting.   
    %    In the geometric term quantum metric term always have minus contribution, quantum metric makes ferro-magnetic/multipole fluctuation. 

    \begin{figure}[tbp]
        %\centering
        %\begin{subfigure}{0.65\linewidth}
            %\centering
            %\caption{Curvature of $\chi^E(\bm{q})$}
            %\includegraphics[width=\linewidth]{ddchiE.png}
          %\end{subfigure}
          %\hfill
       %\begin{subfigure}{0.65\linewidth}
            %\centering
            %\caption{Curvature of $\chi^O(\bm{q})$}   \includegraphics[width=\linewidth]{ddchiO.png}
          %\end{subfigure}
        \includegraphics[width=7cm]{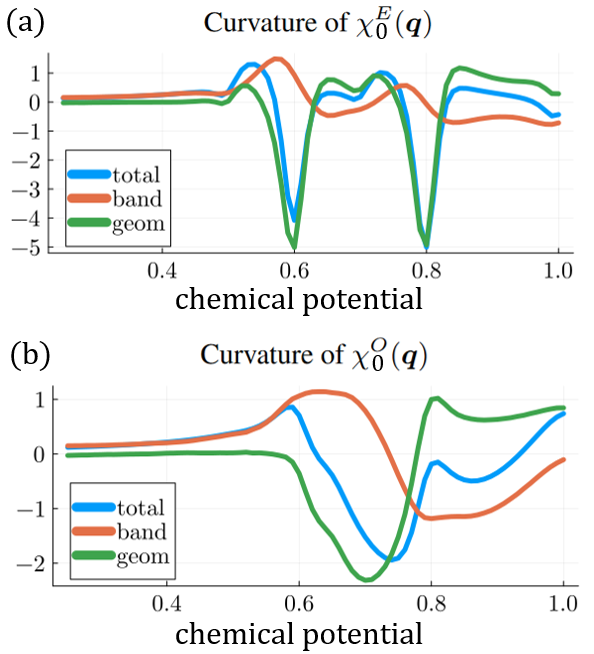}
        \caption{Curvature of (a) spin susceptibility $\chi^{\rm{E}}_0(\bm{q})$ and (b) multipole susceptibility $\chi^{\rm{O}}_0(\bm{q})$ at ${\bm q}=0$, namely, $\partial_{q_x}^2\chi^{\rm{E/O}}_0(\bm{q})\big|_{{\bm q}=0}$. Blue lines represent the total curvature, while red and green lines show the contributions from band dispersion and quantum geometry, respectively.}
        \label{curvature}
    \end{figure}
    
    The above discussions are verified in Fig.~\ref{curvature} which shows the chemical potential dependence of the curvatures of $\chi^{\rm{E/O}}_0(\bm{q})$ at ${\bm q}=0$. The curvature of $\chi^{\rm{E}}_0(\bm{q})$ has sharp dips at $\mu = 0.6$ and $0.8$, while the curvature of $\chi^{\rm{O}}_0(\bm{q})$ shows a broad dip around $\mu=0.7$. %and the dips of $\chi^{\rm{E}}_0(\bm{q})$ are shaper than the dip of $\chi^{\rm{O}}_0(\bm{q})$. 
    Around the dips, the curvatures are mainly determined by the geometric terms and are largely negative because of the quantum metric. In the bilayer Lieb lattice model, the quantum metric is significantly enhanced around the M point because the bands in the mirror-even (mirror-odd) sector degenerate at $\epsilon_n^+({\bm k}_{\rm M})=0.6$ ($\epsilon_n^-({\bm k}_{\rm M})=0.8$).
    The behaviors %of spin and multipole susceptibilities 
    in Fig.~\ref{curvature} are consistent with the discussion in the previous paragraph. Because the quantum metric term in the curvature of spin susceptibility $\chi^{\rm{E}}_0(\bm{q})$ is the Fermi surface term, it is enhanced when the Fermi level lies on the degenerate energy, that is, $\mu=0.6, 0.8$. 
    In contrast, the quantum metric term for the curvature of multipole susceptibility is the inter-band term, and it is enhanced in the relatively wide region, $0.6< \mu < 0.8$. When the temperature is decreased, the dips in spin susceptibility become sharper~\cite{kitamura2024spin}. In contrast, the curvature of multipole susceptibility is not so sensitive to chemical potential and temperature. 

\textit{\HL{Spin and multipole susceptibility}} --- 
    To justify the above arguments regarding the magnetic fluctuations induced by quantum geometry, we also calculate the momentum dependence of the susceptibilities, 
    which are rewritten as %and to illustrate the roles of quantum geometry, we rewrite the bare susceptibilities as %analyze the irreducible susceptibilities for free electrons, which are given by
    \begin{align}
        \chi^{\rm{E/O}}_0(\bm{q}) = 2\int\frac{d\bm{k}}{(2\pi)^2} \sum_{n,m,p} L_{nm}^{p \pm p}(\bm{k},\bm{q})\left( 1 - {D}_{nm}(\bm{k},\bm{q})^2 \right), %NG ok?
    \label{eq:irreducible_susceptibility}
    \end{align}
    with  the quantum distance in the mirror sector $D_{nm}(\bm{k},\bm{q})$ defined as follows 
    \begin{align}
        & \left|\braket{u_n^p(\bm{k}+\bm{q})|u_m^{p^\prime}(\bm{k})}\right|^2 = \delta_{pp^\prime}\left(1-D_{nm}(\bm{k},\bm{q})^2\right).
    \end{align}
    In Eq.~\eqref{eq:irreducible_susceptibility}, $L_{nm}^{p +p}$ ($L_{nm}^{p -p}$) corresponds to the spin (multipole) susceptibility. 
%    The positive sign represents the spin susceptibility and the negative sign represents the multipole susceptibility.
    The quantum distance represents the distance of quantum states in the parameter space, and the second-order derivative of the quantum distance gives the quantum metric~\cite{resta2011insulating}.
    Thus, the effects of quantum geometry are taken into account in the bare susceptibilities through the quantum distance.
We obtain the maximum quantum distance $D=1$ between the even and odd mirror sectors, because the overlap of wave functions with different mirror parity vanishes. %it lead to the maximum quantum distance $D=1$ between the $p=\pm$ mirror sectors. 
Therefore, only the quantum distance in each mirror sector is essential. 
%\YYS{We have defined it by $D_{nm}(\bm{k},\bm{q})$, which is independent of mirror parity and has only band indices $n$ and $m$ (see Appendix~C in Supplemental Material \cite{SM}).}

%In the following, we discuss the quantum-geometric effects on magnetic and multipole orders based on such geometric quantities.   

%    Through mirror eigenvalue independent quantum distance, spin/multipole susceptibility has a quantum geometric effect. As $\bm{q}$ of distance increase, distance also increase. So, quantum distance suppresses antiferro magnetic/multipole susceptibility and makes ferro magnetic/multipole susceptibility.
    To specify the effects of quantum geometry, we compare the susceptibilities to those defined by neglecting the $\bm{q}$-dependence of quantum distances $D_{nm}(\bm{k},\bm{q})$,
    %the momentum dependence of quantum distances, %does not contain quantum geometry by
    \begin{align}
        \chi^{\rm{E/O}}_{0,\rm band}(\bm{q}) &= 2\int\frac{d\bm{k}}{(2\pi)^2} \sum_{n,m,p} L_{nm}^{p \pm p}(\bm{k},\bm{q})\left( 1 - {D}_{nm}(\bm{k},\bm{0})^2 \right), \\
        &= 2\int\frac{d\bm{k}}{(2\pi)^2} \sum_{n,p} L_{nn}^{p \pm p}(\bm{k},\bm{q}).
    \end{align}
    These simplified susceptibilities do not contain quantum-geometric effects because $\bm{q}$-dependence of the quantum distance creates quantum-geometric properties.
    Accordingly, $\chi^{\rm{E/O}}_{0,\rm band}(\bm{q})$ are determined only by band dispersion through the Lindhard function.
    
    \begin{figure}[tbp]
%        \centering
          %\begin{subfigure}{0.45\linewidth}
%            \centering
             %\caption{$\chi^{E}(\bm{q})$}
            %\includegraphics[width=\linewidth]{chi0E.png}
          %\end{subfigure}
          %\hfill
          %\begin{subfigure}{0.45\linewidth}
%            \centering
             %\caption{$\chi^{E}_{\rm band}(\bm{q})$}
            %\includegraphics[width=\linewidth]{chi0NGE.png}
          %\end{subfigure}
            %\hfill
          %\begin{subfigure}{0.45\linewidth}
%            \centering
             %\caption{$\chi^{O}(\bm{q})$}
            %\includegraphics[width=\linewidth]{chi0O.png}
          %\end{subfigure}
            %\hfill
          %\begin{subfigure}{0.45\linewidth}
%            \centering
            %\caption{$\chi^{O}_{\rm band}(\bm{q})$}
          %\includegraphics[width=\linewidth]{chi0NGO.png}
          %\end{subfigure}
        \includegraphics[width=7cm]{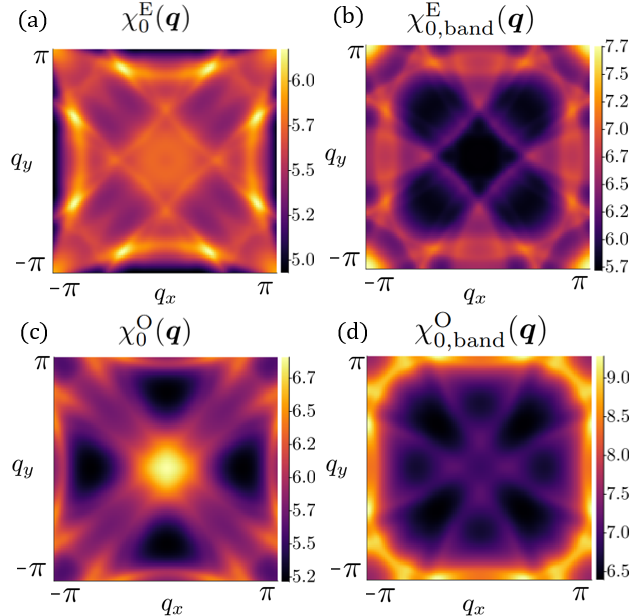}
        \caption{Bare susceptibility for (a) spin fluctuation $\chi^{\rm{E}}_0(\bm{q})$ and (c) multipole fluctuation $\chi^{\rm{O}}_0(\bm{q})$. We show (b) $\chi^{\rm{E}}_{0,\rm band}(\bm{q})$ and (d) $\chi^{\rm{O}}_{0,\rm band}(\bm{q})$ without quantum-geometric effects for comparison. The chemical potential is set to $\mu=0.7$.}
        \label{susceptibility}
    \end{figure}%

Based on the analysis of the curvatures of susceptibilities and the results in Fig.~\ref{curvature}, we can predict the spin and multipole fluctuations, as well as the effects of the quantum metric. %for various chemical potentials.
In particular, complex magnetic correlations are expected to emerge from quantum-geometric effects when one of the spin and multipole fluctuations is dominated by the geometric term and the other is by the band-dispersion term. %band dispersion makes a dominant contribution to the curvature of the spin (multipole) susceptibility, while quantum geometry makes a dominant contribution to the curvature of the multipole (spin) susceptibility.
For such an example, we show the results for the chemical potential $\mu=0.7$ %we obtain the results 
in Fig.~\ref{susceptibility}, where $\chi^{\rm{E/O}}_0(\bm{q})$ are compared to $\chi^{\rm{E/O}}_{0,\rm band}(\bm{q})$.
%Since geometric quantities are known to be enhanced around band degeneracy points, it would be interesting to calculate the spin and multipole susceptibilities by setting the chemical potential close to the degenerate band energies at the M point. For such a chemical potential $\mu=0.7$, we obtain the results in Fig.~\ref{susceptibility}, where $\chi^{\rm{E/O}}_0(\bm{q})$ are compared to $\chi^{\rm{E/O}}_{0,\rm band}(\bm{q})$.
Consistent with Fig.~\ref{curvature},
the quantum-geometric effect significantly modifies the momentum dependence of the odd-parity multipole susceptibility $\chi^{\rm{O}}_0(\bm{q})$, while it has little influence on the spin susceptibility $\chi^{\rm{E}}_0(\bm{q})$. 
In Fig.~\ref{susceptibility}(c), we see a distinguished peak of $\chi^{\rm{O}}_0(\bm{q})$ at ${\bm q}=0$, indicating the ferroic multipole fluctuation. This peak does not appear when we neglect quantum geometry, as shown in Fig.~\ref{susceptibility}(d). The spin susceptibility is also affected by quantum geometry, as we notice by comparing Figs.~\ref{susceptibility}(a) and \ref{susceptibility}(b). However, both $\chi^{\rm{E}}_0(\bm{q})$ and $\chi^{\rm{E}}_{0,\rm band}(\bm{q})$ imply antiferromagnetic fluctuations. The maximum value of multipole susceptibility is larger than that of spin susceptibility.
Combining these results, we conclude that quantum geometry induces ferroic multipole fluctuation rather than ferromagnetic and antiferromagnetic spin fluctuations.

%\YYS{As demonstrated Fig.~\ref{curvature}, the quantum metric favors ferromagnetic fluctuation and ferroic multipole fluctuation in different manners.  Therefore, for $\mu=0.7$, the quantum metric significantly favors the ferroic multipole fluctuation, although the antiferromagnetic fluctuations arising mainly from band dispersion are essentially unaffected by quantum geometry.  The contrasting behaviors of $\chi^{\rm{E}}_0(\bm{q})$ and $\chi^{\rm{O}}_0(\bm{q})$ in Fig.~\ref{susceptibility} are interpreted in this way.} 

    %quantum metric is proportional to $L_{nn}^{pp}(\bm{k},\bm{0}) = -f^{\prime}(\epsilon_n^p(\bm{k}))$ 
    %while in the metric term in the curvature of $\chi^{\rm{O}}_0(\bm{q})$, quantum metric is proportional to $L_{nm}^{p-p}(\bm{k},\bm{0})$. Here $f^{\prime}$ means differential of fermi distribution function. Because of $f^\prime$'s delta function like behavior, metric term of $\chi^{\rm{E}}_0(\bm{q})$ is sensitive to $\mu$, but metric term of $\chi^{\rm{O}}_0(\bm{q})$ is not so sensitive to $\mu$. This difference in $\mu$ dependence create the difference in fluctuation between $\chi^{\rm{E}}_0(\bm{q})$ and $\chi^{\rm{O}}_0(\bm{q})$.
    %need more easy discription

    \begin{figure}[tbp]
        %\centering
          %\begin{subfigure}{0.45\linewidth}
            %\centering
            %\includegraphics[width=\linewidth]{chi0E_mu_04.png}
            %\caption{}
          %\end{subfigure}
          %\hfill
          %\begin{subfigure}{0.45\linewidth}
            %\centering
            %\includegraphics[width=\linewidth]{chi0E_mu_08.png}
            %\caption{}
          %\end{subfigure}
            %\hfill
          %\begin{subfigure}{0.45\linewidth}
            %\centering
            %\includegraphics[width=\linewidth]{chi0O_mu_04.png}
            %\caption{}
          %\end{subfigure}
            %\hfill
          %\begin{subfigure}{0.45\linewidth}
            %\centering
            %\includegraphics[width=\linewidth]{chi0O_mu_08.png}
            %\caption{}
          %\end{subfigure}
        \includegraphics[width=7cm]{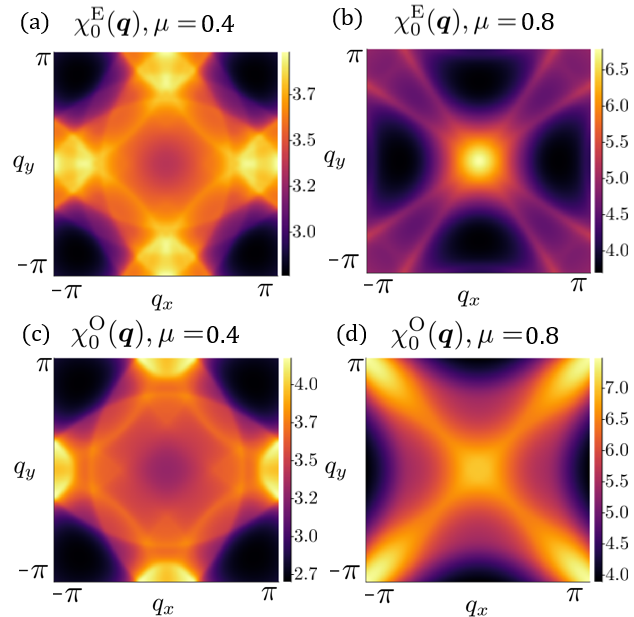}
        \caption{Bare spin susceptibility $\chi^{\rm{E}}_0(\bm{q})$ at (a) $\mu=0.4$ and (b) $\mu=0.8$. Bare multipole susceptibility $\chi^{\rm{O}}_0(\bm{q})$ at (c) $\mu = 0.4$ and (d) $\mu=0.8$.}
        \label{change_mu}
    \end{figure} 
    Figure~\ref{change_mu} shows the susceptibilities $\chi^{\rm{E/O}}_0(\bm{q})$ for $\mu = 0.4$ and $0.8$. First, we expect that quantum geometry is not essential for $\mu = 0.4$, which is lower than the degenerate energies at the M point. %in the mirror-even and mirror-odd sectors. 
    In Figs.~\ref{change_mu}(a) and \ref{change_mu}(c), we see that both $\chi^{\rm{E}}_0(\bm{q})$ and $\chi^{\rm{O}}_0(\bm{q})$ indicate antiferroic spin/multipole fluctuations. The spin and multipole susceptibilities are similar in this case, as is usually expected when the inter-layer coupling is small. Note that $\chi^{\rm{E}}_0(\bm{q}) = \chi^{\rm{O}}_0(\bm{q})$ when the inter-layer hopping is zero. 
    Second, for $\mu = 0.8$, quantum-geometric effects are expected to be significant only for spin susceptibility $\chi^{\rm{E}}_0(\bm{q})$, because the chemical potential lies in the degenerate energy in the mirror-odd sector. 
    This is verified in Fig.~\ref{change_mu}(b), which shows ferromagnetic fluctuation through a distinguished peak in $\chi^{\rm{E}}_0(\bm{q})$ at ${\bm q}=0$. In contrast, multipole fluctuation is rather antiferroic in this case, as shown in Fig.~\ref{change_mu}(d). 
    All of the results are consistent with the contribution of the quantum metric that can induce ferroic fluctuations, implying that quantum-geometric spin and multipole correlations are predictive.  

    %of quantum-geometric magnetism and multipole order 
    %$\chi^{\rm{O}}_0(\bm{q})$ does not have a ferro multipole fluctuation. This is consistent for Fig.\ref{curvature}. 

    \begin{figure}[tbp]
        %\centering
          %\begin{subfigure}{0.45\linewidth}
            %\centering
            %\includegraphics[width=\linewidth]{chiE.png}
            %\caption{}
          %\end{subfigure}
          %\hfill
          %\begin{subfigure}{0.45\linewidth}
            %\centering
            %\includegraphics[width=\linewidth]{chiO.png}
            %\caption{}
          %\end{subfigure}
        \includegraphics[width=7cm]{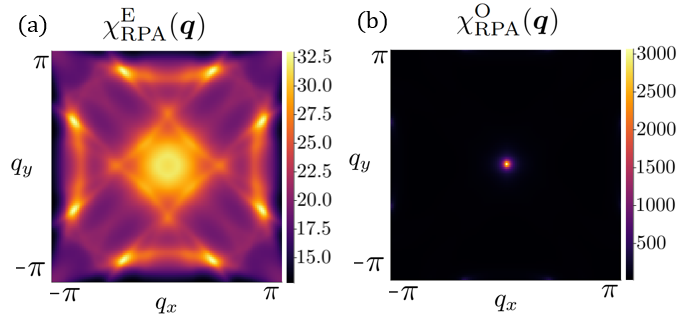}
        \caption{Results of RPA for (a) spin susceptibility $\chi^{\rm{E}}_{\rm RPA}(\bm{q})$ and (b) multipole susceptibility $\chi^{\rm{O}}_{\rm RPA}(\bm{q})$. We set $(U,\mu) = (1.09, 0.7)$.}
        \label{RPA}
    \end{figure}
    
So far, we discussed spin and multipole fluctuations based on the bare susceptibilities. The long-range order that corresponds to the peak of bare susceptibilities can be stabilized by the Coulomb interaction~\cite{Nogaki2024}.
To see this, we calculate the spin susceptibility of even-parity $\chi^{\rm{E}}_{\rm RPA}(\bm{q})$ and the multipole susceptibility of odd-parity $\chi^{\rm{O}}_{\rm RPA}(\bm{q})$ in the interacting system based on the random phase approximation (RPA). The formulation is shown in Section~D in Supplemental Material~\cite{SM}.
Figure~\ref{RPA} shows the RPA susceptibilities $\chi^{\rm{E/O}}_{\rm RPA}(\bm{q})$ obtained for $\mu = 0.7$ and the Hubbard interaction $U=1.09$. 
The multipole susceptibility is divergent at ${\bm q}=0$, indicating the ferroic odd-parity multipole order. 
This is consistent with the fact that the bare ferroic multipole susceptibility is larger than the maximum bare spin susceptibility (see Fig.~\ref{susceptibility}). 
Thus, multipole fluctuations induced by the quantum metric of Bloch electrons condense to ferroic multipole order in interacting systems. 
  %  Habbard interaction increase ferro multipole fluctuation. So, the ferro multipole fluctuation by quantum geometry become phase transition due to RPA interaction.

\textit{Discussions} --- 
For $\mu=0.7$ and $0.8$, the spin and multipole fluctuations exhibit strikingly different momentum dependencies. This behavior is characteristic of systems in which quantum geometry significantly impacts magnetic correlations. 
In the bilayer models, the in-plane and out-of-plane magnetic correlations are anomalously entangled, even when the inter-layer coupling is weak. In contrast, the results for $\mu=0.4$ are typical, with spin and multipole fluctuations nearly equivalent as is expected with a weak inter-layer coupling ($t_{\rm p}=0.1$). In other words, quantum geometry can result in unusual inter-layer spin correlations. Such complex magnetic fluctuations are a signature of quantum-geometric magnetism and can be verified by neutron scattering experiments. Superconductivity mediated by these magnetic fluctuations is an interesting topic for future study.

\textit{Conclusion} ---
In this Letter, we presented the odd-parity multipole magnetism stabilized by quantum geometry. Focusing on the spin multipole defined in multiple sublattices, we have extended recent studies of quantum-geometric ferromagnetism~\cite{kitamura2024spin,kitamura2025quantumgeometricferromagnetismsingular} and shown that the quantum metric suppresses antiferroic multipole correlations and induces ferroic multipole fluctuation. %as well as magnetic fluctuation. 
Coulomb interaction further enhances the multipole fluctuation and leads to the quantum-geometric multipole order.  
    
Our analysis of the curvature of bare susceptibilities has clarified the contrasting effects of the quantum metric on spin and multipole correlations. The quantum metric favors ferromagnetism and ferroic multipole magnetism through a Fermi surface term and an inter-band term, respectively. Therefore, the quantum metric induces ferromagnetism in a narrow region of the chemical potential around the band degeneracy. In contrast, the quantum metric can drive multipole magnetism in a relatively wide range of the chemical potential. %The maximum values of the quantum metric terms are comparable for magnetic and multipole fluctuations. 
    Consequently, quantum geometry, enhanced by topological band crossing, affects multipole correlations more ubiquitously than spin correlations. 
%    The study of the chemical potential dependence of the curvature of the susceptibility shows that the quantum metric term is sensitive to the spin susceptibility, while it is relatively insensitive to the multipole susceptibility. This is due to the shape of the Fermi distribution function appearing in the susceptibility, which is considered an important factor in suppressing fluctuations. Calculations using RPA suggest that the interaction may further enhance the quantum geometry induced ferro multipole fluctuations, leading to a transition to the ordered phase. This implies that quantum geometry also plays a nontrivial role in phase transitions.
%
The results of this work, in particular, the contrasting roles of quantum metric can be applied to various many-body correlations. Quantum geometry can ubiquitously cause unconventional magnetic correlations in multi-sublattice systems. 

\textit{Acknowledgements} ---
    We appreciate helpful discussions with Taisei Kitamura, Chang-geun Oh, and Akito Daido. This work was supported by JSPS
KAKENHI (Grant Numbers JP22H01181, JP22H04933,
JP23K17353, JP23K22452, JP24K21530, JP24H00007, JP25H01249).

\bibliography{bib.bib}

\clearpage
\includepdf[pages=1]{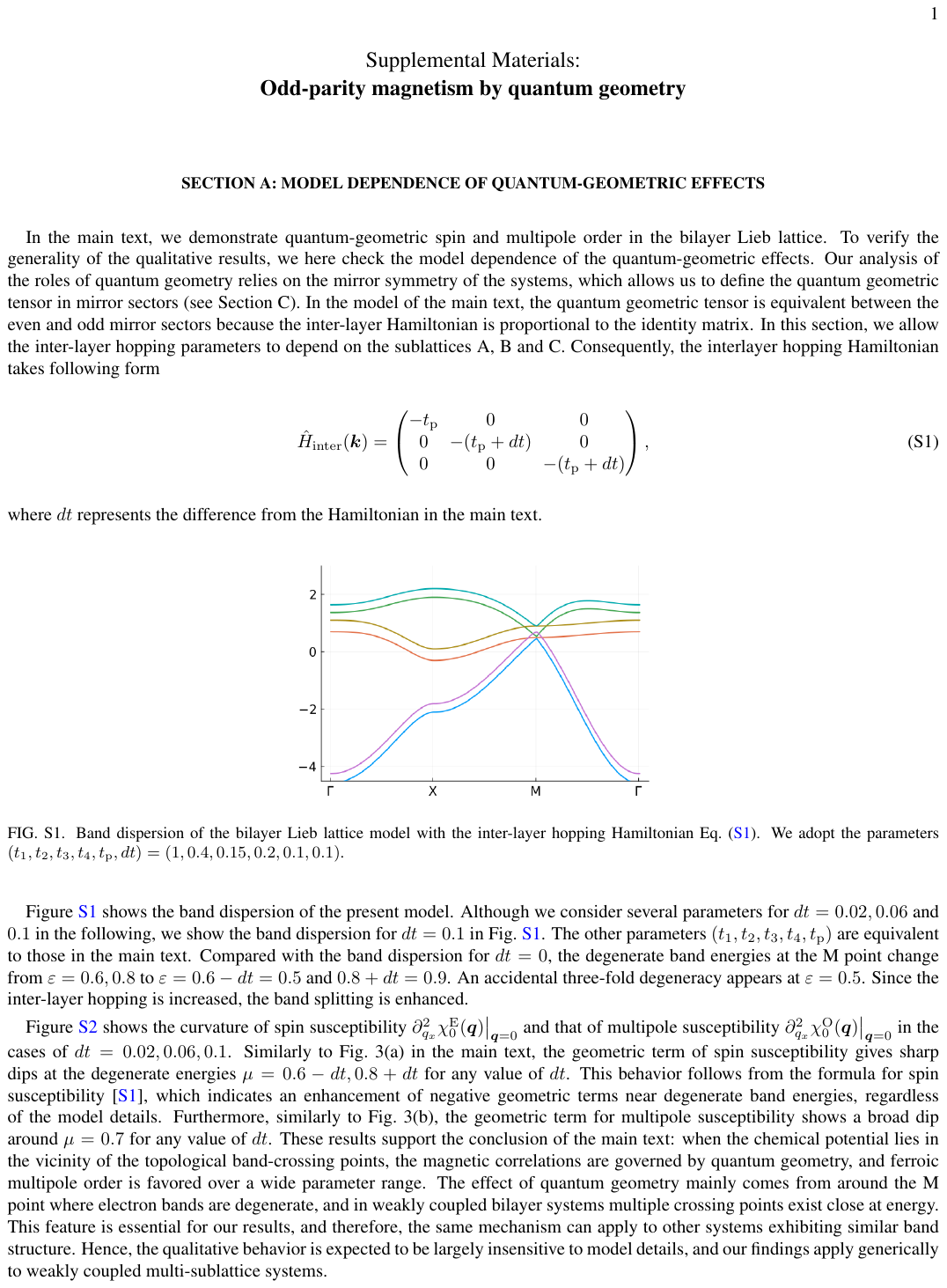}
\clearpage
\includepdf[pages=2]{Supplemantal_Material.pdf}
\clearpage
\includepdf[pages=3]{Supplemantal_Material.pdf}
\clearpage
\includepdf[pages=4]{Supplemantal_Material.pdf}
\clearpage
\includepdf[pages=5]{Supplemantal_Material.pdf}

\end{document}